\documentclass[journal]{IEEEtran}
\usepackage{algpseudocode}
\usepackage{makecell}
\usepackage{array}
\usepackage{ragged2e}
\newcolumntype{J}[1]{>{\justifying\arraybackslash}p{#1}}

\usepackage{amsmath,amssymb,amsfonts}
\usepackage{graphicx}

    \usepackage[left = 1.8 cm, right = 1.8 cm, top = 1.8 cm, bottom = 1.8 cm]{geometry}

\usepackage{enumitem}    

\usepackage[english]{babel}
\usepackage[utf8]{inputenc}
 \usepackage[ruled,vlined]{algorithm2e}
 \usepackage{bm}
 \usepackage{caption} 

 \usepackage{cite}

\ifCLASSINFOpdf
\else
\fi
\usepackage{amsmath,amssymb,amsfonts}

\usepackage{graphicx}
\graphicspath{%
    {converted_graphics/}
    {/}
}

  \makeatletter
\newcommand*{\rom}[1]{\expandafter\@slowromancap\romannumeral #1@}
\makeatother

\usepackage{xcolor}
  
\begin{document}

\title{Trustworthy AI-Driven Dynamic Hybrid RIS: Joint Optimization and Reward Poisoning-Resilient Control in 
Cognitive MISO Networks\\
}

\author{Deemah~H.~Tashman,~\IEEEmembership{Member, IEEE}, 
        and Soumaya~Cherkaoui,~\IEEEmembership{Senior Member, IEEE}%
\thanks{D.~H.~Tashman and S.~Cherkaoui are with the LINCS Laboratory, 
Department of Computer and Software Engineering, Polytechnique Montréal, 
Montréal, QC, Canada, H3T 1J4 (e-mail: \{deemah.tashman, soumaya.cherkaoui\}@polymtl.ca). 
LINCS Lab website: https://lincslab.ca/en/}%
}

 \maketitle

\begin{abstract}
Cognitive radio networks (CRNs) are a key mechanism for alleviating spectrum scarcity by enabling secondary users (SUs) to opportunistically access licensed frequency bands without harmful interference to primary users (PUs). To address unreliable direct SU links and energy constraints common in next-generation wireless networks, this work introduces an adaptive, energy-aware hybrid reconfigurable intelligent surface (RIS) for underlay multiple-input single-output (MISO) CRNs. Distinct from prior approaches relying on static RIS architectures, our proposed RIS dynamically alternates between passive and active operation modes in real time according to harvested energy availability. We also model our scenario under practical hardware impairments and cascaded fading channels. We formulate and solve a joint transmit beamforming and RIS phase optimization problem via the soft actor-critic (SAC) deep reinforcement learning (DRL) method, leveraging its robustness in continuous and highly dynamic environments. Notably, we conduct the first systematic study of reward poisoning attacks on DRL agents in RIS-enhanced CRNs, and propose a lightweight, real-time defense based on reward clipping and statistical anomaly filtering. Numerical results demonstrate that the SAC-based approach consistently outperforms established DRL baselines, and that the dynamic hybrid RIS strikes a superior trade-off between throughput and energy consumption compared to fully passive and fully active alternatives. We further show the effectiveness of our defense in maintaining SU performance even under adversarial conditions. Our results advance the practical and secure deployment of RIS-assisted CRNs, and highlight crucial design insights for energy-constrained wireless systems. 

\end{abstract}


\IEEEpeerreviewmaketitle

\section{Introduction}

\IEEEPARstart{E}{nriched}    immersive experiences, enhanced ubiquitous coverage, and advanced networked intelligence, including seamless human-machine collaboration, are expected to be supported by the evolution toward sixth-generation (6G) networks expected in 2030 \cite{nagarajdemystifying}. This necessitates the integration of cognitive radio networks (CRNs), as they offer the dynamic spectrum management and adaptability necessary to accommodate the heightened demand for bandwidth and diverse applications that are envisioned by 6G networks.
CRN offers different access modes to facilitate secondary users (SUs), who are unlicensed users, in accessing the bands assigned to primary users (PUs), who are   known as licensed users \cite{9348134,11059714}. When SUs attempt to access the frequency band via the underlay mode, they must ensure that their transmission power does not surpass the interference threshold that the PU receiver can tolerate \cite{9237455,10368012}. Additionally, in order to address the growing need for reliable connectivity in 6G CRN networks, it is necessary to incorporate additional technologies alongside CRN. Nevertheless, the incorporation of active components, such as cooperative relaying, necessitates energy consumption, which presents a challenge in energy-limited devices \cite{10188924} and lacks built-in intelligence, limiting their ability to autonomously adapt to changing environments.

To resolve the aforementioned challenges, reconfigurable intelligent surfaces (RISs) can be considered. RISs comprise multiple tiny components, which intelligently direct signals to boost communication coverage and strengthen signals for the designated receivers \cite{10584518,10921906}. RIS operates in   passive and active modes. In passive RIS, a central control unit is responsible for adjusting the phase of the elements according to real-time channel information \cite{10319408}.   The lack of power amplification for signals in this type results in marginal energy consumption, rendering it both affordable and energy-efficient \cite{10584518}. Nonetheless, passive RISs are disadvantaged by the multiplicative fading effect,  leading to substantial performance degradation \cite{10778216}. Consequently, active RISs have been implemented to address this problem   by  modifying the phase of the RIS elements and amplifying the power of the signal, augmenting communication coverage and reliability. However,  active RIS results in significant power consumption and thermal noise, which can impair performance, and add expense of hardware implementation. Hence, hybrid passive-active RISs have  been utilized to merge the advantages of both modes while alleviating their disadvantages to balance cost and performance. However, fixed hybrid RIS designs, where certain elements are designed to be active and the remainder are passive, lack adaptability to   energy conditions, limiting their efficiency in scenarios like cognitive Internet of things (IoT).

Deep reinforcement learning (DRL) has enabled substantial progress in optimizing CRNs \cite{10182973,10592377,10794361} and ideal passive  RIS systems \cite{10361836,9766179,10039119,10257607,10319408,wei2024dynamic}.  Active RISs, although promising for overcoming multiplicative fading, have only recently been explored for CRN reliability using DRL  with a fixed amplification gain in \cite{10636057}.   Hybrid RISs have been employed in satellite security and cellular contexts with DRL-based optimization in \cite{10483088,10810415,10807185}, but not, to date, for reliability or dynamic energy awareness in CRNs.  Despite the evident positive outcomes achieved by DRL techniques, cyber-physical security for DRL, and specifically reward poisoning attacks, is a growing concern that yields detrimental training results.  In CRNs, reward manipulation can cause spectrum misuse and degrade performance, threatening network reliability and user trust. Up until now, a limited  number of theoretically defense methods exist \cite{10679438,banihashem2023robust,nika2023online,liu2025certified}, and they are often computationally intensive or require access to multiple environments. This leaves a gap in lightweight, practical defense techniques, which this work attempts to address.

\vspace{0.5em}
\noindent

\vspace{0.5em}

\noindent\textbf{Contributions:}
Up to the authors knowledge, no previous work examined a dynamic energy-aware hybrid RIS for enhancing the reliability of CRNs, nor systematically defended  DRL-based RIS agents against reward poisoning. In contrast to previous works that presume ideal passive RIS reflection or fixed-gain active RIS, our paper introduces a dynamic hybrid RIS architecture that incorporates non-ideal reflection, realistic amplification models, and energy harvesting (EH) limitations; and studies lightweight DRL defense mechanisms in adversarial settings.  Given this, the main contributions are summarized as follows:
{\color{black}\begin{itemize}
  \item \textbf{Dynamic Energy-Aware Hybrid RIS \& Joint Optimization:} We propose a dynamic hybrid RIS that switches between passive/active operation based on the harvested energy, and jointly optimizes SU transmit beamforming and RIS coefficients in an underlay multiple-input single-output (MISO)-CRN. The RIS is assumed to harvest energy from a dedicated power beacon, providing an energy-aware amplification approach where the reflected signal amplitudes and RIS mode switching depend on the actual energy budget.
  
  \item \textbf{Robust Modeling \& Performance Evaluation:} We develop a realistic system model including phase-dependent (non-ideal) reflection, energy-aware active amplification with amplifier noise, and cascaded Rayleigh channels. We evaluate our scenario using the   SUs' average rate  and we study the impact of several key parameters. We also provide an energy-consumption model comparing active vs. hybrid modes, showing the achieved energy savings.  A trade-off between energy efficiency and performance is also conducted. 
  
  \item \textbf{DRL Control under Constraints:} We design a soft actor–critic (SAC) controller that maximizes SU rates  under realistic, non-ideal channel and device assumptions. A comparison between the SAC approach and other DRL techniques is also provided. 
  
  \item \textbf{Reward Poisoning Defense:} For the first time, reward poisoning attacks on DRL agents in RIS-assisted CRNs are investigated and a lightweight defense mechanism (statistical filtering and reward clipping) is demonstrated.
\end{itemize}}

The rest of the paper is organized as follows; the related works are given in section \rom{2}. The system model is presented in 
 section \rom{3}. The SUs' rate maximization problem is provided in section \rom{4}. We present the DRL approach in section \rom{5}. Moreover, the poisoning attacks and the defense mechanism are proposed in section \rom{6}. The results  are provided in section \rom{7}, while a future work direction is given in section \rom{8}. Finally, conclusions are illustrated in section \rom{9}.

\section{Related Works}

DRL approaches have shown exceptional performance in replacing complex conventional optimization problems performed by RIS control unit in   CRNs \cite{10361836,10622458}. For instance, in \cite{9766179}, the transmit power of the SUs and the reflection array of the passive RIS were jointly optimized. This optimization was performed while ensuring that the quality of the received signals at the PUs met certain constraints. The optimization problem was addressed using the deep deterministic policy gradient (DDPG) and SAC methodologies. Moreover, a strategy that utilizes a passive RIS to facilitate dynamic spectrum sharing was introduced in \cite{10039119}. The RIS assists the SUs in dynamically accessing the spectrum by leveraging the eavesdropped automatic repeat request (ARQ) feedback from the PUs. The authors employed DRL methodologies to achieve the most efficient spectrum-sharing scheme. Furthermore, the authors in \cite{10257607} demonstrated the utilization of passive RIS to enhance the security of the SUs in a CRN. To achieve the highest level of confidentiality, the subchannel assignment, transmit beamforming of the SU transmitter, and RIS reflection coefficients were optimized using multiple deep Q networks and DDPG techniques. In \cite{10319408}, a passive multi-RIS to enhance spectrum sharing and ensure secure transmission performance was proposed. The utilization of double deep Q-networks and SAC algorithms aimed to maximize the secrecy rate of the SUs by optimizing the pairings of the RIS, the assignment of subchannels, and the transmit beamforming of the SUs network. Additionally, in \cite{wei2024dynamic}, the authors used a passive RIS to assist the communication of unmanned aerial vehicle (UAV)-assisted wideband CRNs. An optimization problem was proposed to jointly optimize the trajectories of the primary and secondary UAVs, power allocation at the secondary UAVs, and reflection coefficients of RIS to maximize the achievable rate of SUs via DRL techniques. 

So far, only a single study has explored the utilization of DRL methodologies to enhance the reliability of CRNs through the implementation of active RISs. In \cite{10636057}, an underlay cognitive multicast communication system supported by active RISs has been assumed. The authors suggested optimizing the beamforming matrix at the base station and the active RIS to maximize the minimal signal-to-interference-plus-noise ratio (SINR) using DDPG and twin delayed deep deterministic policy gradient (TD3) DRL algorithms.

Hybrid RISs have lately emerged as a promising technology, drawing increased attention from researchers due to their capacity to balance cost and performance. For example, in \cite{10483088}, the authors examined the security of a satellite downlink communication system utilizing a hybrid RIS. To optimize the system's worst-case secrecy rate, the authors jointly optimized the satellite design and RIS beamforming employing a deep post-decision state–deterministic policy gradient (DPDS-DPG) DRL method. Comparable assessments were performed for a MISO satellite downlink communication system in \cite{10810415} utilizing a DDPG-based approach. To mitigate security threats, the study in \cite{10807185} employed the DDPG DRL approach to optimize a hybrid RIS for security maximization. However, dynamic hybrid RISs have yet to be explored for enhancing reliability or adapting to energy variations in CRNs.

\par In wireless systems, such as CRNs,   reward manipulation can lead to spectrum misuse and degraded performance, posing serious risks to network reliability and user trust. Therefore, designing lightweight and effective defense mechanisms against reward manipulation is essential to ensure the reliability of DRL techniques. Recent research has shown that reward poisoning attacks may significantly degrade the performance of reinforcement learning (RL) by manipulating only a small fraction of the reward signals in both online and offline learning settings \cite{10679438,cai2023reward,xu2022efficient}. Black-box attacks such as U2 \cite{zhang2020adaptive} and RankPoison \cite{xu2022efficient} show that attackers can still interfere with the learning process even without access to the RL model's internals. Recent studies have begun addressing reward poisoning attacks in DRL. Bouhaddi and Adi \cite{10679438} proposed a multi-environment training and variance-based detection strategy. Moreover, in \cite{banihashem2023robust}, the authors introduced a provably robust optimization framework. Nika et al. \cite{nika2023online} modeled  the defense as a bandit game with sublinear-regret algorithms, while Liu et al. \cite{liu2025certified} provided certified offline RL defenses using differential privacy.    While a limited number of theoretically robust methods exist for mitigation, they are often computationally intensive or require access to multiple environments. This produces a gap in lightweight, practical defense mechanisms.  Fast, low-latency defenses such as reward clipping combined with statistical mean-variance filtering have been little studied—a gap this work addresses. Table~\ref{tab:related-work} summarizes the key differences between our proposed work and closely related studies.  

\begin{table}[!h]
\centering
\caption{Comparison of Closely Related Works}
\label{tab:related-work}
\resizebox{\columnwidth}{!}{%
\begin{tabular}{|l|c|c|c|c|c|}
\hline
\textbf{Ref.} & \textbf{RIS Mode} & \makecell{\textbf{Energy} \\ \textbf{Adaptation}} & \makecell{\textbf{Channel/} \\ \textbf{Hardware}} & \makecell{\textbf{RL} \\ \textbf{Security}} & \textbf{DRL} \\
\hline
\cite{10483088, 10810415, 10807185} & Fixed hybrid   & No  & \makecell{Ideal/ideal}      & No     & DDPG/DPG \\
\cite{10319408, 9766179, 10039119, 10257607, wei2024dynamic} & Passive        & No  & \makecell{Ideal/ideal}   & No     & DDPG/SAC \\
\cite{10636057}   & Active         & No  & \makecell{Ideal/ideal}            & No     & DDPG/TD3  \\
\textbf{Ours}     & Dynamic hybrid & Yes & \makecell{Cascaded \\ /non-ideal} & Yes    & SAC      \\
\hline
\end{tabular}%
}
\end{table}

\vspace{0.5em}
\noindent

\vspace{0.5em}

\noindent\textbf{Novelty Summary:}
Compared to the above, this work is the first to design and analyze a dynamic, energy-adaptive hybrid RIS for CRNs under practical energy harvesting constraints and hardware impairments, with joint DRL optimization. This paper also presents a systematic defense against reward poisoning attacks using simple and efficient statistical filtering mechanisms.

\IEEEpeerreviewmaketitle

{\color{black}The following are the notations employed in this paper:  Vectors are denoted by bold lowercase letters (e.g., $\mathbf{x}$), matrices by bold capital letters (e.g., $\mathbf{H_s}$), $(\cdot)^T$ signifies the transpose operation,  $(\cdot)^\mathcal{H}$ is the Hermitian transpose (conjugate transpose) operator,    $\mathrm{Tr}(\cdot)$ signifies the trace, $|\cdot|$ represents the magnitude of a complex number, $\mathbb{E}[\cdot]$ indicates the expectation operator, and $\mathrm{diag}(\mathbf{x})$ refers to the diagonal matrix constructed from vector $\mathbf{x}$. Moreover, the main symbols  employed throughout the paper and their definition are summarized in Table \ref{tablesymbols} for further clarity.}

 \section{System Model} 
\begin{figure}[hb]
  \centering
  \includegraphics[width=1.0\linewidth]{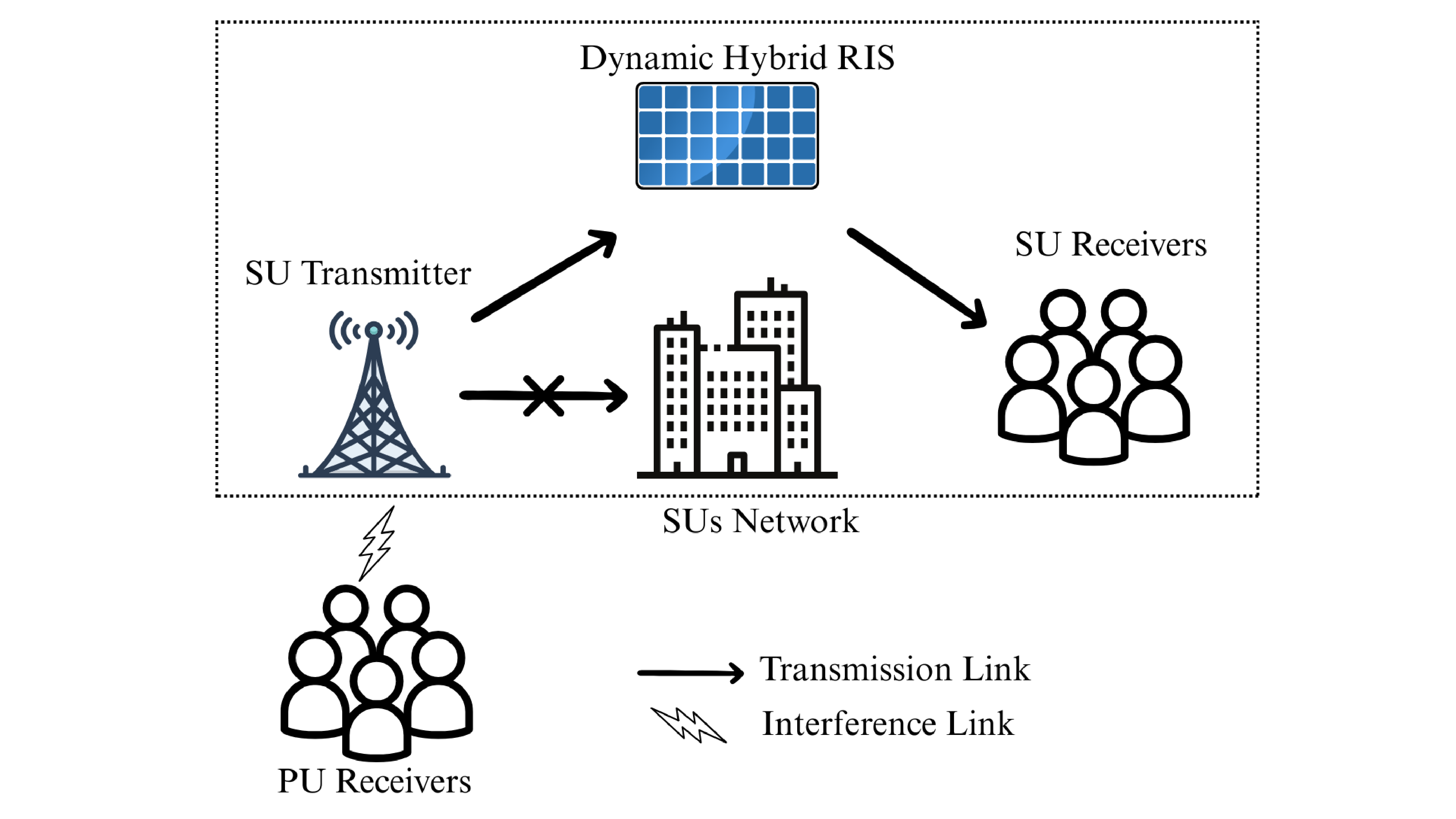}
  \caption{The system model.}
  \label{sys1}
\end{figure}

Figure \ref{sys1} depicts a MISO underlay CRN, where there is an SU transmitter equipped with $A$ antennas, aiming to establish communication with $B$ SU receivers, each equipped with a single antenna. It is presumed that the direct connection between the SUs is not feasible given the occurrence of significant signal blockage \cite{bae2024overview}. Therefore, we examine an RIS that is installed on a building and has $R$ reflecting elements. Additionally, there is a micro-controller that aids in directing the received signals towards the desired SU receivers.  Moreover, there are $W$ PU receivers that have a designated interference threshold that must be adhered to by the SU transmitter. The channel matrix between the SU transmitter and the RIS is represented by the  complex-valued matrix $\mathbf{H_s} \in \mathbb{C}^{R \times A}$. Similarly, the channel vector between the RIS and the $B$ SU receivers is  a complex-valued vector indicated by $\mathbf{h_{b}}\in  \mathbb{C}^{R \times 1}$, and $b$ is an index that ranges from 1 to $B$. In this scenario, we assume that the RIS operates in hybrid mode. That is, it can   switch between passive and active modes depending on the available harvested energy, as will be elaborated in the subsequent subsections.

\begin{table}[h]
    \small
    \centering
    \caption{List of Main Symbols}\label{tablesymbols}
   \begin{tabular}{|c|J{6.2cm}|}  
    \hline
    \textbf{Symbol} & \textbf{Definition}  \\
    \hline
    $A$ & Number of antennas at SU Transmitter  \\
    $B$ & Number of SU Receivers \\
    $R$ & Number of RIS reflecting elements \\
    $W$ & Number of PU receivers \\
    $\mathbf{H_s}$ &  SU transmitter and RIS link channel matrix   \\
    $\mathbf{h_b}$ &  RIS and   SU receivers channel vector  \\
    $\mathbf{H_p}$ &   SU transmitter and the PU receivers channel \\ & gain  \\
     $\mathbf{h_{\text{PB}}}$ & Power beacon and RIS channel gain   \\
    $\Phi$ & Diagonal phase shift matrix at the RIS \\
    $\Psi$ & RIS reflection matrix in the active mode \\
    $\epsilon_r$ & Phase shift applied by the RIS element \\
    $\beta(\epsilon_r)$ & Reflection amplitude of the passive RIS ele-\\ & ments \\  
    $\beta_m$ & Minimum passive RIS reflection amplitude  \\
    $m$, $l$ & Hardware-dependent RIS parameters control-\\ &ling the   reflection amplitude degradation with \\ &phase shift \\
    $\mathbf{G}$ & Beamforming matrix at SU transmitter \\
    $\kappa_s,\kappa_b, \kappa_p$ & Cascade level for $\mathbf{H_s}$, $\mathbf{h_b}$, and $\mathbf{H_p}$\\  
    $\alpha_{\text{scaled}}$ & Amplification gain of active RIS elements \\
    $E_{\text{total}}$ & Total energy harvested \\
    $\tau$ & Minimal energy required to operate in active \\ & mode \\
    $I$    &  PU receivers interference threshold \\
    $P_t$   &  Maximum SU transmitter available power  \\
    $\eta$ &   Energy conversion efficiency \\
    $A_T$ &   Reward threshold \\
    $S_C$ &   Scaling factor \\
    \hline  
    \end{tabular}
\end{table}

\subsection{Passive RIS}
Passive RISs are employed to steer signals toward the SU receivers, hence optimizing the transmission data rate for SUs, while offering low energy consumption and reduced hardware complexity.  The majority of  prior research has considered ideal reflection by the RIS such that the signal power is lossless from each RIS reflection element. However, assuming the hardware impairments and  similar to \cite{10283517} and \cite{9115725}, we presume that the RIS follows the  phase-dependent amplitude model, such that the reflection coefficient   depends on the phase shift.

Let $\mathbf{\Phi}=\operatorname{diag}[\phi_1,\phi_2,\cdots,\phi_R]$ be the diagonal phase shift matrix at the reflecting RIS. The entries of $\mathbf{\Phi}$ are given as $\phi_r=\beta(\epsilon_r)e^{j\epsilon_r}$,   such that  $\epsilon_r\in[0,2\pi)$ is the phase shift applied by the RIS element. $\beta(\epsilon_r)$ represents the reflection amplitude of the $r^{th}$ element of the RIS and it is given as \cite{10283517}
\begin{IEEEeqnarray}{lCr}   \label{passive non ideal}
 \beta(\epsilon_r)=\left(1-\beta_m\right) \left(\frac{\sin\left(\epsilon_r-l\right)+1}{2}\right)^{f} +\beta_m,
\end{IEEEeqnarray}
\noindent where $\beta_m\in[0,1]$ represents the minimum value of the reflection amplitude for each element of the RIS. $l \geq 0$ and $m \geq 0$ are both parameters dependent on the hardware implementation of the RIS. Given this, the received signal at the  SU receiver $b$ is given by
\begin{IEEEeqnarray}{lCr} \label{yb} 
y_b= \mathbf{h_b^T \Phi H_s G x} +n_b,
\end{IEEEeqnarray}
\noindent where  $\mathbf{G} \in \mathbb{C}^{(A\times B)}$ is the beamforming matrix at the SU transmitter, $\mathbf{x}$ is a data stream column vector of dimension $B\times 1$ to be transmitted to all the $B$ SU receivers, and $n_b$ is the additive-white-Gaussian-noise (AWGN) with zero mean and variance $\sigma_b^2$.

Prior CRNs-enhanced RIS studies have presumed that the channels are characterized by standard fading distributions, such as   Rayleigh fading model. However, classical channels have been proven insufficient for accurately representing the connections in mobile devices and rich scattering areas, as demonstrated in \cite{9408651}. Experimental studies have demonstrated that cascaded channels are better suited than non-cascaded channels for representing the transmission of signals \cite{10466378,9612017,9094381}.   Cascaded channels, also known as multiplicative channels, are formed when many signals are multiplied together after being reflected off scatters or objects, resulting in the received signal \cite{10464644}. Thus, we assume that the SUs' network links, i.e., $\mathbf{H_s}$ and $\mathbf{h_b}$ follow the cascaded Rayleigh fading model.    Given this, let $\xi=\prod_{j=1}^{\kappa_i} X_j$,   for $\xi \in \{\mathbf{H_s,h_b}\}$, $i \in\{s,b\}$, and with  $X_j$ following the Rayleigh fading model.  $\kappa_i$ represents the cascade level for the link $i$, such that $\kappa_s$ is the cascade level of the link $\mathbf{H_s}$ and $\kappa_b$  is the cascade level of the link $\mathbf{h_b}$.

Given $(\ref{yb})$, the SINR at each SU receiver is given as
\begin{IEEEeqnarray}{lCr}  \label{snr}
\lambda_b^{p}=\frac{\left| \mathbf{h_b^T \Phi H_s g_b }  \right|^2}{ 
 \sum_{r,r\neq b}^B  \left|\mathbf{h_b^T \Phi H_s g_r} \right|^2 +\sigma_b^2  },
\end{IEEEeqnarray}
\noindent where $\mathbf{g_b}$ is the $b^{th}$ column vector of $\mathbf{G}$, and the denominator in $(\ref{snr})$ denotes the co-channel interference plus noise variance $(\sigma_b^2)$. To evaluate the system performance, we use the ergodic sum rate as
\begin{IEEEeqnarray}{lCr}  \label{rate}
q  \left(\mathbf{ G,\Phi,h_b,H_s}\right) = \sum_{b=1}^B D_b^{p},
\end{IEEEeqnarray}
\noindent where $D_b^{p}$ is the data rate at the $b^{th}$ SU receiver, such that $D_b^{p}=\log_2 \left(1+\lambda_b^{p}\right)$.

 \subsection{Active RIS}
Despite the advantages of passive RIS in efficiently steering signals toward the destination with negligible power consumption, its performance can degrade in high-noise environments or under severe fading, where signal amplification becomes essential. This is addressed by active RISs, where each reflecting element not only applies a phase shift but also amplifies the incoming signal. However, active RIS introduces significant power consumption, as amplification requires additional energy. To overcome this challenge, we propose an energy-aware amplification approach in which the RIS is wirelessly powered by a dedicated power beacon \cite{9860773}, enabling it to adapt its operation based on the harvested energy. This energy aware behavior introduces a practical (non-ideal in terms of performance) reflection technique, where the amplification gain varies according to the available power at each RIS unit. The entries of the RIS reflection matrix $\boldsymbol{\Phi}$ are defined as $\phi_r = \alpha_r \, e^{j \epsilon_r},$ where $\alpha_r > 1$ is the energy-dependent amplification gain. The following subsections describe our proposed EH methodology and its incorporation into the active RIS model.

 \subsubsection{Energy Harvesting Model}
To overcome the energy consumption problem in the active RIS, we assume that the RIS is equipped with an EH capability \cite{9926102}. That is, the RIS harvests energy from a   dedicated power beacon  as in \cite{9860773}, such that the harvested energy at the $r$-th RIS element is expressed as 
\begin{equation}
    E_r = \eta   \left| \mathbf{h}_{{\text{PB}}}^{(r)} \right|^2   P_{\text{PB}} T,
\end{equation}
\noindent where  $\eta$ denotes the energy conversion efficiency, $P_{\text{PB}}$ is the transmission power of the power beacon,  $\mathbf{h}_{\text{PB}}^{(r)}$ represents the   channel  gain from the power beacon to the $r$-th RIS element, modeled as Rayleigh fading, and $T$ denotes duration of the EH phase.

\subsubsection{Active Mode Gain Calculation}
For a practical and energy-efficient active RIS, we propose a new formulation that constrains the amplification gain based on the harvested energy, rather than assuming a constant gain as in ideal active RIS reflection, thereby reflecting realistic hardware restrictions.  This prevents unnecessary amplification at low energy levels while facilitating robust signal amplification when sufficient power is available.   
The amplification gain at each RIS element is determined by adjusting a base gain in accordance with the harvested energy.  Consequently, the energy-aware scaling is defined by
\begin{equation} \label{scale}
    \alpha_{\text{scaled}}  = \alpha_{\min} + (\alpha_{\max}  - \alpha_{\min})  f,
\end{equation}
\noindent where $\alpha_{\min} > 1$ denotes the minimum amplification gain, guaranteeing that the RIS maintains a degree of signal amplification even in conditions of insufficient energy (no element is entirely deactivated). Moreover, $f$ is the subsequent energy ratio   to demonstrate the extent of amplification permissible per RIS element, and it is given by
\begin{equation}
   f = \frac{E_{total}/R}{E_{\max}},
\end{equation}
\noindent where $E_{max}$ is the energy required to achieve a maximum amplification gain $(\alpha_{max})$. $ E_{\text{total}} = \sum_{r=1}^{R} E_r$ is the total energy harvested across all $R$ RIS elements.   This energy-aware ratio is uniformly applied to adjust the gain  across all RIS elements, in accordance with the shared energy budget.
Furthermore, to guarantee that the amplification gain does not exceed the hardware-imposed upper limit $(\alpha_{\max})$, we implement a hard constraint defined as 
\begin{equation}
    \alpha_{\text{scaled}} = \min(\alpha_{\text{scaled}}, \alpha_{\max}).
\end{equation}
\noindent This higher clipping ensures hardware safety and reliability by preventing over-amplification caused by abnormal energy values or unusual signal configurations.  Given this, the received signal at the $b$th SU receiver under the active RIS model is expressed as 
\begin{equation}
    y_b^{a} = \mathbf{h}_b^{\mathrm{T}} \Psi \mathbf{H}_s \mathbf{G} \mathbf{x} + 
    \mathbf{h}_b^{\mathrm{T}} \Psi \mathbf{w}_r + n_a,
\end{equation}
\noindent where \(\mathbf{\Psi} =  \alpha_{\text{scaled}} \operatorname{diag} \left(  e^{j \epsilon_1},   e^{j \epsilon_2}, \dots,  e^{j \epsilon_R} \right) \), and \( \mathbf{w}_r \sim \mathcal{CN}(\mathbf{0}, \sigma_r^2 \mathbf{I}_R) \) denotes the thermal noise introduced by the RIS amplifiers, with \( \mathbf{I}_R \) representing the identity matrix of size \( R \times R \). It is worth mentioning that for practical feasibility, we consider a  uniform amplification gain across all RIS elements based on the total harvested energy, i.e., $\alpha_r = \alpha_{\text{scaled}},\ \forall r$. {\color{black} Moreover, note that $\mathbf{\Psi}$ denotes the modified RIS reflection matrix in the active mode, reformulated from the general phase-shift matrix  $\boldsymbol{\Phi}$ to include the energy-dependent amplification factor $\alpha_{\text{scaled}}$.}

\noindent Similar to (\ref{snr}), the SINR at the $b$th SU receiver when the RIS is in active mode is expressed as
\begin{equation}
    \lambda_b^{a} = 
    \frac{\left| \mathbf{h}_b^{\mathrm{T}} \Psi \mathbf{H}_s \mathbf{g}_b \right|^2}
    {\sum\limits_{r \neq b} \left| \mathbf{h}_b^{\mathrm{T}} \Psi \mathbf{H}_s \mathbf{g}_r \right|^2 + 
    \sigma_r^2 \left\| \mathbf{h}_b^{\mathrm{T}} \Psi \right\|^2 + \sigma_a^2},
\end{equation}
where $\sigma_r^2 \left\| \mathbf{h}_b^{\mathrm{T}} \Psi \right\|^2$ accounts for the thermal noise introduced by the RIS amplifiers, and $\sigma_a^2$ is the noise variance at the receiver. The corresponding ergodic sum rate is defined as
\begin{equation}
    z(\mathbf{G}, \Psi, \mathbf{h}_b, \mathbf{H}_s) = 
    \sum_{b=1}^{B} A_b^{a},   
\end{equation}
\noindent where $A_b^{a}= \log_2(1 + \lambda_b^{a})$.
 \subsection{Dynamic Hybrid RIS}
In the dynamic hybrid RIS mode, we propose an energy threshold to determine whether the elements should function as passive (reflecting only) or active (reflecting and amplifying) through the threshold $\tau$. $\tau$ denotes the minimal energy required for the RIS to operate in the active mode.  Hence, the decision concerning the RIS mode is presented as
\begin{equation} \label{compare}
\text{Mode} =
\begin{cases}
\text{Passive}, & \text{if } E_{\text{total}} < \tau \\
\text{Active}, & \text{if } E_{\text{total}} \geq \tau
\end{cases} .
\end{equation}
 \noindent In (\ref{compare}), if the available energy is insufficient to support active amplification, the RIS defaults to passive mode to maintain functionality without violating power constraints. 

{\color{black} In the proposed hybrid RIS model, it is assumed that each RIS element is equipped with a switching mechanism that enables it to dynamically toggle between passive and active modes, as given in \cite{wang2023hybrid}. Conceptually, this can be realized through a dual-path architecture where each element includes a passive and an active load controlled via RF switches (e.g., PIN-diode or MOSFET-based). To support energy-aware operation, each element can integrate an RF-to-DC rectifier, local energy storage (e.g., a supercapacitor or thin-film battery), and a low-power power-management integrated circuit (IC) to monitor the harvested energy. The controller can then apply a threshold rule based on the stored energy to activate the amplification path when sufficient energy is available or revert to the passive path otherwise. While our work remains theoretical, this assumption is supported by recent hardware prototypes \cite{wang2023hybrid} that integrate tunable phase shifters and low-noise amplifiers, suggesting a feasible path toward practical large-scale hybrid RIS deployment in future wireless systems.}

\subsection{Energy Consumption for the Hybrid RIS}
Given  $P_{\text{passive}}$  as the per-element control power required to operate a passive RIS element  when the RIS operates in the passive mode, the total energy consumed across all $R$ elements is expressed as $R   P_{\text{passive}}$. Additionally, when the RIS operates in the active mode, the total energy consumption includes both amplification and control overhead for each element \cite{10480438}. The total energy consumed by the hybrid RIS operation is given as
\begin{equation}
    E_{\text{hybrid}} =
    \begin{cases}
        R   P_{\text{passive}}, & \text{if } E_{\text{total}} < \tau \\
        R \left( \alpha_{\text{scaled}}   P_{\text{amp}} + P_{\text{ctrl}} \right), & \text{if } E_{\text{total}} \geq \tau
    \end{cases},
\end{equation}
\noindent where   $P_{\text{amp}}$ is the power consumed per unit amplification, and $P_{\text{ctrl}}$ is the control circuit power per element \cite{10480438}. It is worth mentioning that the power consumption of passive RIS elements is typically very low and nearly negligible. However, for completeness and fairness in comparison, we still include this minimal consumption in the overall energy calculation of the hybrid RIS.

\subsection{PUs' Interference Constraint}

To avoid interfering with the PUs' transmission, we need to make sure that the transmission power of the SU transmitter is maintained below the interference threshold tolerable at the PU receivers $(W)$. Hence, this constraint is given as \cite{8203951}
\begin{IEEEeqnarray}{lCr}  \label{const}
\mathbf{E} [tr\{\mathbf{Gx}(\mathbf{Gx})^\mathcal{H}\}] \leq \min \{P_t,\frac{I}{\max \left({g_{sp1},{g_{sp2},\cdots,{g_{spW}}}}\right)}\}, \nonumber \\
\end{IEEEeqnarray}
\noindent where  $P_t$ is the maximum transmission power available at the SU transmitter  and $I$ is the interference threshold tolerable at PU receivers.  $g_{spi}$, for $i \in\{1,2,\cdots, W\}$, is the channel power gain of the link between the SU transmitter and each PU receiver $i$. The MISO channel gain between the SU transmitter and the PUs is represented by $\mathbf{H_p}\in \mathbb{C}^{A \times W}$, such that the entries of this matrix, i.e., the channel gain between SU-Tx and each PU follows the cascaded Rayleigh fading model with a cascade level of $\kappa_p$. Similar to \cite{9500621}, we assume that the PU receivers are located far away from the building where the RIS is mounted, and thus cannot be impacted by its transmissions.

\section{Optimizing Beamforming and RIS Matrices}
 Our primary goal is to maximize the sum data transmission rate of the SUs' network by optimizing the beamforming matrix employed by the SU transmitter and the reflecting phases of the RIS. Similar to \cite{10622614}, we do not optimize the amplitude of the RIS reflection coefficients, as implementing independent control of both amplitude and phase shifts is costly in practice.  Our optimization problem is carried out while ensuring that the transmission power does not disrupt the communication of the PUs, and that the hybrid RIS operates in the active mode only when sufficient energy is harvested. Given this, the optimization problem is formulated as
\hspace*{-\IEEEiednormlabelsep}%
\begin{IEEEeqnarray}{rl}   \label{opti-prob}
\mathcal{P}: \quad & \underset{\mathbf{j}, \mathbf{G}}{\text{max}} \;\; 
    \mu\left(\mathbf{G}, \mathbf{j}, \mathbf{h}_b, \mathbf{H}_s\right) \\
\text{s.t.} \quad & tr\left\{\mathbf{G G}^\mathcal{H} \right\} \leq \min \left\{ P_t, \right. \nonumber \\
& \left. \frac{I}{\max(g_{sp1}, g_{sp2}, \cdots, g_{spW})} \right\}, \IEEEyessubnumber \label{firstrate1} \\
& |\epsilon_r| \in [0, 2\pi), \quad \forall r = 1, 2, \cdots, R, \IEEEyessubnumber \label{secondcon} \\
& E_{\text{total}} \geq \tau, \IEEEyessubnumber \label{thirdcon}
\end{IEEEeqnarray}

\noindent for $\mu \in \{q, z\}$ and $\mathbf{j} \in \{\mathbf{\Phi}, \mathbf{\Psi}\}$.

\section{DRL-based Solution to Maximize SUs' Reliability}
This section provides a description of the key components of the DRL agent and provides the method  used to solve the optimization problem given in (\ref{opti-prob}).  Given that the state at each time slot is dependent on the state of the preceding time slot, the problem described in (\ref{opti-prob}) meets the criteria for the Markov property. Therefore, it may be represented as a model-free Markov decision process (MDP) \cite{10646359}. Hence, this model is  represented by the following components:

\begin{itemize}
    \item State space $(S)$: The state space consists of several components, including the transmission power of the SU transmitter at the $t^{th}$ time slot, the maximum interference threshold permitted at the PU receivers, the channel matrices $(\mathbf{H_s^t, h_b^t} \forall b$, \text{and}  $\mathbf{H_p^t})$, and the action executed during the previous time step. Hence, the state space at time step $t$ is defined as

    \begin{IEEEeqnarray}{lcr} 
    S_t=\{P_t^t, I^t, \mathbf{H_s^t, h_b^t, H_p^t, G^{(t-1)}, \Phi^{(t-1)}\Psi^{(t-1)}}\}. \nonumber \\
\end{IEEEeqnarray}

\item Action space $(Ac)$: It consists of the DRL agent's decisions about the transmit beamforming matrix and the RIS phase shift matrix at each time step. Hence, the action space at each time step is given as    
    \begin{IEEEeqnarray}{lcr} 
    A_{c,t}=\{\mathbf{G^{t}, \Phi^{t}, \Psi^t}\}.
\end{IEEEeqnarray}

\item Reward $(R_d)$: The received reward at each time step is the sum rate of the SUs achieved through interactions with the environment \cite{9318243,9729992,9524882,10078092,9685236,9839316,9999295,10243611} and it is divided according to the mode of operation, i.e., passive or active. Hence, the reward is given as    
\begin{equation} \label{rewardfold}
 R_{d,t}  =
\begin{cases}
q\left(\mathbf{G}^{t}, \mathbf{\Phi^{t}}, \mathbf{h}_b, \mathbf{H}_s \right)   \\
z\left(\mathbf{G}^{t}, \mathbf{\Psi^{t}}, \mathbf{h}_b, \mathbf{H}_s \right) - w_t   \max(0, \tau - E_{\text{total}}) ,
\end{cases}
\end{equation}
\noindent where $w_t$ is the penalty weight. The first reward in (\ref{rewardfold}) is the one received if the RIS is functioning in the passive mode. To accommodate for the energy constraint in (\ref{thirdcon}), the agent will be penalized with an energy-dependent term when operating in the active mode but insufficient energy is harvested. Specifically, the second case in (\ref{rewardfold}) corresponds to the active RIS mode, where the sum rate $z(\cdot)$ is reduced by a penalty that increases as the harvested energy $E_{\text{total}}$ falls below the minimum threshold $\tau$. This formulation incentivizes the agent to not only optimize the transmission performance but also maintain energy-efficient RIS configurations that are sustainable under realistic EH conditions.

\end{itemize}
In order to address the optimization problem presented in (\ref{opti-prob}), we use the SAC  DRL technique.  SAC is renowned for effectively managing continuous states and action spaces, which aligns with the requirements of our environment.

The Q-value function in  RL is used to assess the impact of selecting an action on the anticipated future reward under a given policy $\pi$. Therefore, the Q-function is expressed as \cite{9110869}
    \begin{IEEEeqnarray}{lcr} 
   Q_\pi\left(s_t,a_t\right)= E_\pi\left(R_t|s_t=s,a_t=a\right),
\end{IEEEeqnarray}
\noindent where $R_t=\sum_{w=0}^{\infty} \gamma^w r^{t+w+1}$, $s_t$ is the state at time $t$, $a_t$ is the action taken at time $t$, $r$ is the instant reward, and  $\gamma$ is the discount factor, ranges from $0$ to $1$, and indicates the extent to which the agent prioritizes long-term rewards over immediate ones. The provided Q-function satisfies the well recognized Bellman equation, expressed as
    \begin{IEEEeqnarray}{lcr} 
   \label{bellman}Q_\pi\left(s_t,a_t\right)= E_\pi\left(r^{t+1}|s_t=s,a_t=a\right) \\ \nonumber +\gamma \sum_{s'}p_s\left(\sum_{a'} \pi(s',a')Q(s',a')\right),
\end{IEEEeqnarray}

\noindent with $p_s$ being the probability of moving from state $s$ to state $s'$ after taking action $a$. The optimal Q-function, i.e., $Q^{\ast}(s_t,a_t)$ can be found from (\ref{bellman}) as 
\begin{IEEEeqnarray}{lcr}\label{Qopt}
Q^\ast\left(s_t,a_t\right)= r^{t+1}\left(s_t=s,a_t,\pi=\pi^\ast \right)\\ \nonumber+\gamma\sum_{s'}p_s \max_{a'} Q^\ast(s',a'),
\end{IEEEeqnarray}
\noindent where $\pi^\ast$ is the optimal policy.

In RL, the Bellman equation is solved using a recursive approach to get the optimal Q-function given in (\ref{Qopt}). The method of updating the Q-function is defined as follows
    \begin{IEEEeqnarray}{lcr} 
   Q^\ast\left(s_t,a_t\right) \leftarrow (1-\Upsilon_q) Q^\ast(s_t,a_t)\\ \nonumber+\Upsilon_q\left(r^{t+1}+\gamma\max_{a'}Q_\pi(s_{t+1},a'\right) .
\end{IEEEeqnarray}
\noindent The learning rate, denoted as $\Upsilon_q$, controls the magnitude of the parameters updates during training. It plays a crucial role in finding a balance between stability and learning speed.  When the action and state spaces are continuous and large, updating the Q-value at each time step becomes intricate. Hence, deep neural networks (DNNs) can be effectively employed to approximate the Q-function. The integration of DNNs in RL leads to the notion of DRL. In DRL, the Q-value function is determined by the DNN parameters $\theta$, which include the weights and biases as
    \begin{IEEEeqnarray}{lcr} 
   Q\left(s_t,a_t\right) = Q\left(\theta|s_t,a_t\right) .
\end{IEEEeqnarray}
\noindent The optimum Q-value function may be achieved by updating the DNN parameter $\theta$ using stochastic optimization techniques. The parameter $\theta$ is updated as
    \begin{IEEEeqnarray}{lcr} 
   \theta_{t+1}=\theta_t-\Upsilon_\theta \nabla_\theta \mathcal{L}(\theta),
\end{IEEEeqnarray}
 \noindent where $\Upsilon_\theta$ represents the learning rate used to update the parameter $\theta$, while $\nabla_\theta$ denotes the gradient of the loss function $\mathcal{L}(\theta)$ with respect to $\theta$. The loss function is computed as the difference between the estimated Q-value and the target Q-value. Given this, the loss function is defined as
  \begin{IEEEeqnarray}{l}
    \mathcal{L}(\theta) = \bigg( r^{t+1} + \gamma \max_{a'} Q\left(\theta^{trg} \mid s_{t+1}, a'\right)  \nonumber \\
    \hspace{18pt} - Q\left(\theta^{trn} \mid s_t, a_t\right) \bigg)^2,
\end{IEEEeqnarray}

\normalsize
\noindent where $Q\left(\theta^{trg}|s_{t+1},a'\right)$ is the target neural network and $Q\left(\theta^{trn}|s_t,a_t\right)$ is the training neural network. 

SAC is an off-policy approach that is advantageous for resolving our problem for two primary reasons. Initially, we assume that both the state and action spaces are continuous, and SAC algorithm is well-known for its proficiency in performing successfully in such situations. Furthermore,   SAC  has proven to surpass previous DRL methods \cite{10283517,9500202,9110869}.  Three networks are present in SAC: two Q-networks (critic networks) and an actor network. The actor network is used to select actions based on a state, and it represents the policy. The actor network follows a stochastic policy, which means that it generates a probability distribution over actions rather than a singular deterministic action. The two Q-networks are used to evaluate the expected return of their input, which are the state-action pairs.

\noindent In SAC approach, the Q-networks are jointly trained as
   \begin{IEEEeqnarray}{lcr} 
  U=r+\gamma\min_{i=1,2} Q_{\theta_{i}'}\left(s',a'\right)-e_s \log \left(a'|s'\right) , 
\end{IEEEeqnarray} 
   \begin{IEEEeqnarray}{lcr} 
   Z(\theta_i)=\frac{1}{M}\|U-Q_{\theta_{i}} (s,a)\|_2^2, 
\end{IEEEeqnarray} 
   \begin{IEEEeqnarray}{lcr} 
    \theta_i  \leftarrow \theta_i-  \Upsilon_s \nabla_{\theta_{i}} Z(\theta_i) , 
\end{IEEEeqnarray}

\noindent where $\theta_i$ are the corresponding parameters to the $i^{th}$ Q-network, $M$ represents the mini-batch size that is sampled from the experience relay buffer, and $\Upsilon_s$ is the learning rate. $Z(\theta_i)$ denotes the loss, whereas $\nabla_{\theta_{i}} Z(\theta_i)$ represents the gradient of the loss with respect to $\theta_i$.
The entropy parameter, denoted as $e_s$, governs the level of exploration in the SAC algorithm. Greater values of $e_s$ equate to more exploration, motivating the agent to investigate a broader spectrum of actions.  Although a DRL method with a deterministic policy may be used, it necessitates the inclusion of additive noise in order to facilitate exploration. On the other hand, in SAC, entropy regularization modifies the value of $e_s$ according to the current policy's knowledge, using a mechanism called automatic entropy tuning. This approach adjusts the value of $e_s$ in order to achieve a balance between exploration and exploitation, by aligning the policy's entropy with a target entropy value. Therefore, the adaptive method of SAC enhances its effectiveness in addressing complex problems such as transmit beamforming and RIS phase shift design. This is achieved by ensuring thorough exploration without relying on external noise, thus optimizing the learning process.

The policy network receives state vectors as input and produces action vectors as output. The loss function for this network is defined as
   \begin{IEEEeqnarray}{lcr} 
  Z(\zeta) = \frac{1}{M} \sum_{j=1}^{M} \left( e_s \log \pi_{\zeta}(a_{j}|s_{j}) - \min_{i=1,2} Q_{\theta_{i}}(s_{j}, a_{j}) \right), \nonumber \\ 
\end{IEEEeqnarray}
\noindent where $\pi_{\zeta} (a_{j}|s_{j}) $   is the   stochastic policy function that outputs the probability of taking action $a_j$ given the  state $s_j$   parameterized by $\zeta$. The gradient of the loss $(\nabla_\zeta Z(\zeta))$ is then computed by the stochastic policy gradient approach and is utilized for updating the parameters through the gradient decent as 
   \begin{IEEEeqnarray}{lcr} 
   \zeta \leftarrow \zeta - \Upsilon_p \nabla_\zeta Z(\zeta).
\end{IEEEeqnarray}
\noindent It is worth mentioning that we adopt a non-episodic approach, as the considered tasks lack a defined terminal state, and thus the agent is trained using average-reward RL to ensure stable learning and consistent long-term performance. {\color{black} The proposed   SAC procedure is described in Algorithm~\ref{alg:sac_ris}.}

\begin{algorithm}[h]
{\color{black}
\caption{Soft Actor–Critic (SAC) with Energy-Aware RIS for Maximizing SUs' Reliability}
\label{alg:sac_ris}
\KwIn{Environment $\mathcal{E}$, replay buffer $\mathcal{D}$, learning rates, discount factor $\gamma$, energy threshold $\tau$, and total time steps $T_{\text{max}}$}
\KwOut{Trained policy $\pi$ and optimized beamforming and RIS configurations}

\For{each random seed}{
    Initialize policy $\pi$; training critics; target critics; and replay buffer $\mathcal{D}$\;
    Reset the environment and obtain the initial state $s_0$\;
    
    \For{$t = 1$ \KwTo $T_{\text{max}}$}{
        Select action $a_t \sim \pi(s_t)$, where $a_t$ represents both the transmit beamforming matrix $\mathbf{G}^{t}$ and RIS phase shift matrix $(\boldsymbol{\Phi}^{t}, \boldsymbol{\Psi}^{t})$\;
        
        Execute $a_t$ in $\mathcal{E}$ and observe next state $s_{t+1}$, reward $r_t$, and harvested energy $E_{\text{total}}$\;

        \eIf{$E_{\text{total}} < \tau$}{
            Set RIS to \textbf{passive} mode (reflection only)\;
        }{
            Set RIS to \textbf{active} mode and scale amplification according to $E_{\text{total}}$\;
        }

        Store transition $(s_t, a_t, r_t, s_{t+1})$ in $\mathcal{D}$\;

        Sample a mini-batch from $\mathcal{D}$ and update the SAC components as follows:
        \begin{itemize}
            \item \textbf{Critic update:} minimize the Bellman loss \\ between predicted Q-network values and \\ target values from target network;
            \item \textbf{Actor update:} minimize the entropy-\\regularized policy loss using $\pi_{\zeta}$ and $Q_{\theta_i}$;
            \item \textbf{Temperature update:} adjust $e_s$ automatically\\ to maintain target entropy;
            \item \textbf{Target update:} softly update targets.
        \end{itemize}

        Update $s_t \leftarrow s_{t+1}$ and repeat until convergence\;
    }
}}
\end{algorithm}


  {\color{black} \section{Reward Poisoning Attack \& Defense in Deep Reinforcement Learning}

 Poisoning attacks usually take place during the training phase. In RL, poisoning attacks can occur by manipulating the agent's rewards  to mislead the training process of a victim. For instance, a reward poisoning attack can be conducted by adding the perturbation  to the reward so that the victim receives an adversarial sample  as its immediate reward. Similar to \cite{zhang2020adaptive}, we assume that  the attacker has knowledge of the environment MDP and the DRL agent’s Q-learning algorithm. Given this, we assume that the attacker performs an adaptive reward attack. That is, the attacker performs a reward manipulation by inverting or scaling the reward whenever the agent is performing well in terms of the reward. Specifically, if   the estimated Q-value was larger than a threshold $(A_T)$, i.e., if the agent's recent average rewards indicate high performance, the attacker flips the reward's sign (invert attack) or multiplies the reward with a scaling factor $(S_C < 1)$. As a result, the agent is misled into learning that beneficial actions are harmful, which can significantly degrade the learning process over time.

\begin{algorithm}[h]
{\color{black}
\caption{DRL–SAC with Energy-Aware RIS and Reward Defense}

\KwIn{Network parameters, replay buffer, reward statistics (mean, standard deviation)}

\KwOut{Trained policy and RIS configuration for each random seed}

\BlankLine

\For{each random seed}{
    Initialize SAC network parameters, replay buffer\; warm-start reward statistics  (mean $(\vartheta)$ and standard deviation ($\upsilon$)) using the first  $w_p$ nominal rewards collected before any poisoning occurs\;

    \Repeat{convergence}{
        Observe current state $S_t$\;

        Select actions: beamforming vector and RIS phase shift\;

        Compute raw reward $r_t$ and clip: $\hat{r}_t = \text{clip}(r_t, r_\text{min}, r_\text{max})$;

\eIf{$| \hat{r}_t - \vartheta | \le \chi \cdot \upsilon$}{
    Accept $\hat{r}_t$ as $r_t^{\text{final}}$, store transition $(S_t, a_t, r_t^{\text{final}}, S_{t+1})$ in replay buffer;
    Update running mean $(\vartheta)$ and standard deviation ($\upsilon$) using accepted $r_t^{\text{final}}$;
}{
    Discard reward (do not store poisoned transition);
}

        Update Q-networks, policy, and entropy temperature using SAC update rules\;

        \eIf{$E_{\text{total}} < \tau$}{
            Set RIS to \textbf{passive} mode\;
        }{
            Set RIS to \textbf{active} mode and adjust amplification accordingly\;
        }
    }
}
\label{SACalgorithm}}
\end{algorithm}

 {\color{black} To mitigate this attack, we use a statistical filtering and reward clipping filter. Given this,   the reward statistics (mean and standard deviation) are initialized using a short warm-up phase that collects the first $w_p$ nominal, unperturbed rewards to establish a stable baseline at the beginning of training. Then, the observed reward is clipped to lie within a predefined range $[r_{min},r_{max}]$ to bound extreme poisoned values. The mean and standard deviation-based filter is then used to check the clipped reward. If it falls within a particular range of the running mean of clean rewards (e.g., within $\chi$ standard deviations), the reward is  acceptable; otherwise, it is discarded. Specifically, the mean and standard deviation are computed over the most recent accepted rewards in the replay buffer, allowing the defense to dynamically adapt to gradual environment changes while filtering out abnormal deviations.}  This outlier filtering helps detect and eliminate abnormal reward values, such as those that result from manipulation, before they are used to update the agent's policy. This way the agent can maintain stable learning dynamics even under adversarial interference. The final reward used for the agent's training is expressed as 
\begin{IEEEeqnarray}{lcr}
r_t^{\text{final}} =
\left\{\begin{array}{ll}
\hat{r}_t, & \quad \text{if } |\hat{r}_t - \vartheta| \leq \chi \cdot \upsilon \\
\text{discarded}, & \quad \text{otherwise}
\end{array},
\right.   
\end{IEEEeqnarray}
\noindent where $\hat{r}_t = \text{clip}(r_t, r_{\min}, r_{\max})$, and $\vartheta$ and $\upsilon$ are the mean and standard deviation of previously observed clean rewards, respectively. {\color{black} The proposed reward defense mechanism is integrated into the SAC procedure, as presented in Algorithm~\ref{SACalgorithm}.}

It is worth mentioning that while prior work addresses advanced RL attacks such as policy/observation poisoning, we focus on reward manipulation for practical reasons: CRNs can experience hijacked feedback links or misbehaving nodes, making reward signals inherently vulnerable. The proposed statistical clipping defense, though simple, offers minimal computational overhead and is compatible with real-time, resource-constrained deployments—a necessity in practical RIS-assisted CRNs. Given this, in Table \ref{tab:defense-comparison}, we provide a comparison between our defense mechanism  and existing approaches in the literature, showing that our proposed method offers minimal computational cost and is simple to integrate into existing DRL techniques. 

\begin{table}[!h]
\centering
\caption{Comparison of Reward Poisoning Defense Strategies}
\label{tab:defense-comparison}
\resizebox{\columnwidth}{!}{
\begin{tabular}{|l|c|c|c|}
\hline
\textbf{Defense Type} & \makecell{\textbf{Computational} \\ \textbf{Cost}}  & \makecell{\textbf{Ease of} \\ \textbf{Integration}}  & \textbf{Ref.} \\
\hline
Clipping + Statistics         & Minimal       & Very easy     & \textbf{Ours} \\
Multi-env Training   & High          & Moderate      & \cite{10679438} \\
Provable Robust Optimization  & Very High     & Difficult     & \cite{banihashem2023robust} \\
Bandit Game Defense           & Moderate      & Moderate      & \cite{nika2023online} \\
Certified Offline RL       & Very High     & Difficult     & \cite{liu2025certified} \\
\hline
\end{tabular}}
\end{table}

\section{Simulation Results}
In this section, we test the effectiveness of the SAC approach in tackling our optimization problem. Moreover, for the SAC approach,   the tuned   hyperparameter settings were influenced by the ones  outlined in \cite{10283517}.  Unless otherwise specified in the caption, the parameters are set to the following values:    $A=2$, $B=2$, $R=4$, $W=2$, $\kappa_s=\kappa_b=4$, $\kappa_p=1$, $w_t=0.1$, $I=10$ dB, $P_t=10$ dB, $\eta=0.9$, $P_{\text{PB}}=10$ Watt (W), $\alpha_{min}=1.2$, $\alpha_{max}=2$,  $e_s=0.2$,  $\beta_m=0.6$,   $\tau=50$ Joules, $T=1$,   $l=0$, $r_{min}=-2$, $r_{max}=2$, $\chi=2$, $w_p=10$,  and $m=1.5$. {\color{black} It is worth mentioning that the DRL parameters (learning rate ($\Upsilon_p=\Upsilon_s=10^{-3})$, batch size $(M=16)$, and discount factor $(\gamma=1)$) were chosen following standard DRL benchmarks for non-episodic approaches to ensure training stability in continuous action spaces \cite{10283517}. Each experiment was repeated with 10 random seeds, which is the DRL benchmarking standards defined in \cite{henderson2018deep}, and results were averaged to capture consistent learning behavior.  }

{\color{black} \subsection{Impact of RIS elements $(R)$, cascade elements $(\kappa)$, RIS minimum reflection amplitude $(\beta_m)$, and maximum transmission power available at SU-Tx $(P_t)$}}
{\color{black}This subsection examines the influence of several crucial parameters on the SUs' rate, as seen in Figure \ref{two}, Figure  \ref{three}, and  Figure \ref{seven}.   Specifically,  we explore the impact of the RIS elements $(R)$, cascade elements $(\kappa)$, RIS minimum reflection amplitude $(\beta_m)$, and maximum transmission power available at SU-Tx $(P_t)$.}

Figure \ref{two} illustrates the sum rate at the SUs for various values of the number of RIS reflecting elements denoted as $R$. It is evident that after the first training stages are completed, the agent acquires the optimal beamforming and RIS reflection phases that result in the highest possible rate. This is seen via the convergence of the results towards a certain value.  Furthermore, it is observed that when the value of $R$ grows, the performance of the SUs' communication enhances and becomes more consistent.  With more elements, the RIS is able to better regulate the propagation of signals, resulting in improved signal strength. In addition, the increase in multi-paths resulting from multiple RIS components significantly enhances the total data transmission rate of the SUs.

  \begin{figure}  [t] 
  \centering
  \includegraphics[width=0.9\linewidth]{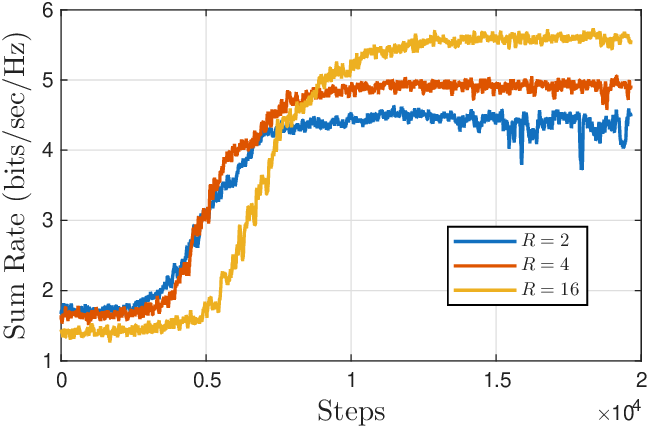}
  \caption{SUs' sum rate for different RIS reflecting elements $(R)$.  $A=4$, $B=4$,   $P_{t}=30$ dB, $I=20$ dB,  and   $\kappa_s=1$, $\kappa_b=1$. }
  \label{two}
\end{figure}

  \begin{figure}  [t] 
  \centering
  \includegraphics[width=0.9\linewidth]{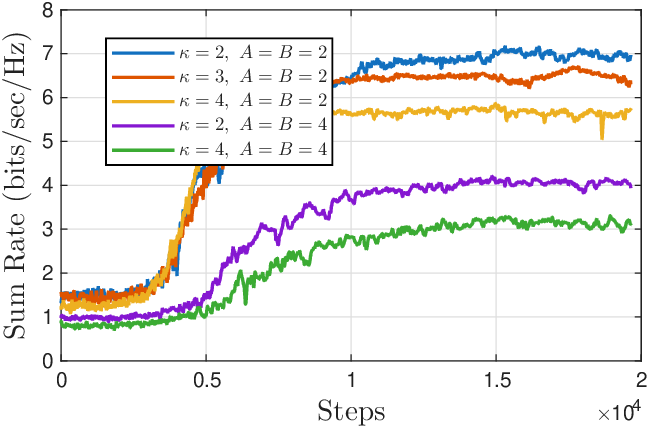}
  \caption{SUs' sum rate for different cascade levels $(\kappa)$, SU transmitter antennas ($A$), and SU receivers $(B)$.   $R=30$,  $P_{t}=30$ dB, $I=20$ dB, and   $\kappa_s=\kappa_b=\kappa$.}
  \label{three}
\end{figure}

 \begin{figure}  [t] 
  \centering
  \includegraphics[width=1.0\linewidth]{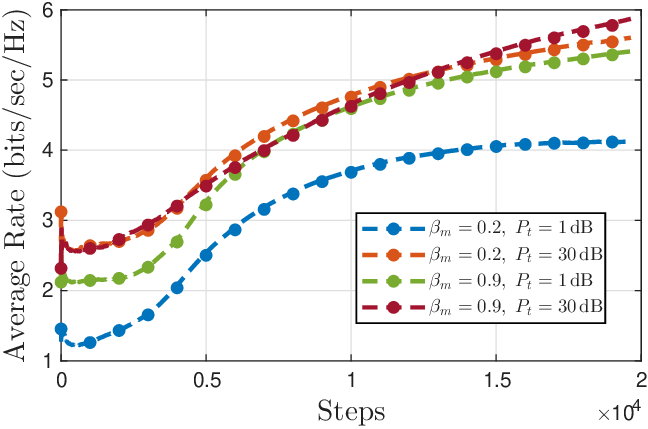}
  \caption{SUs' average rate (reward) for different values of the RIS minimum reflection amplitude $(\beta_m)$ and the maximum transmission power available at the SU transmitter  $(P_t)$.     $\kappa_s=\kappa_b=1$, and $I=20$ dB. }
  \label{seven}
\end{figure}

   \begin{figure}  [b] 
  \centering
  \includegraphics[width=1.0\linewidth]{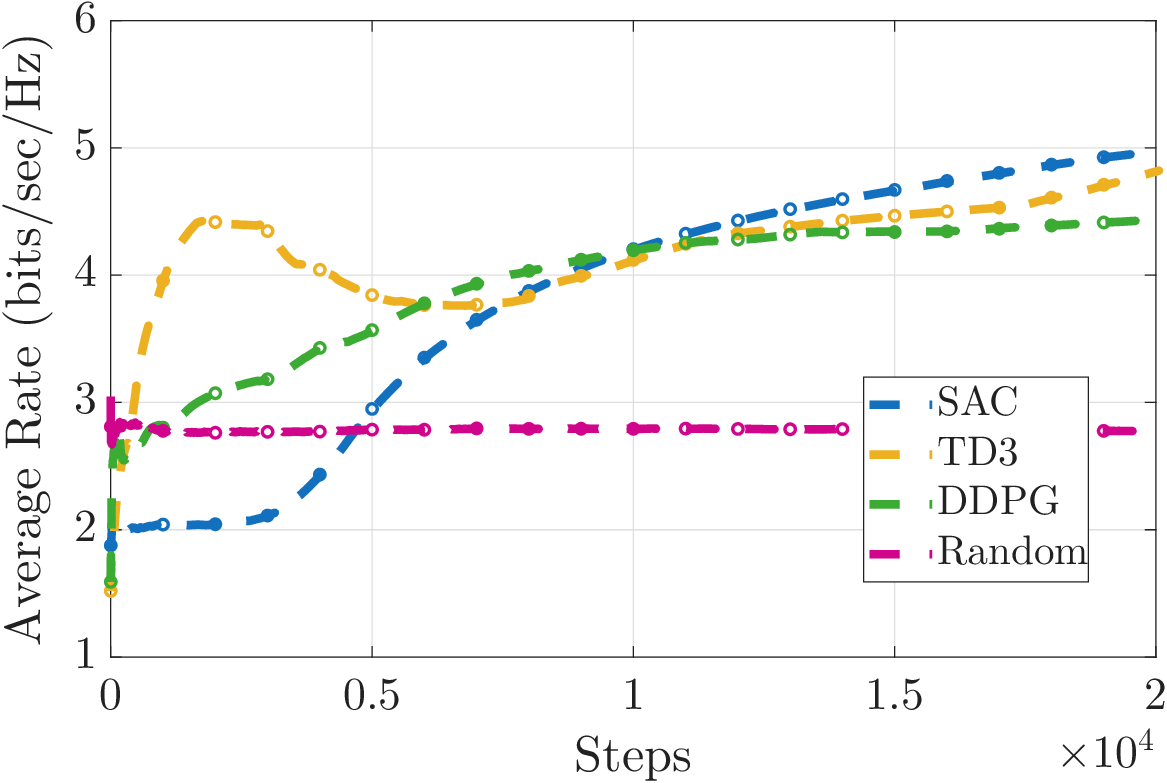}
  \caption{ {\color{black} A comparison between the SUs' average rate for the SAC,   TD3, DDPG, and Random    policies.   $P_{t}=1$ dB, $\kappa_s=\kappa_b=1$, $I=1$ dB,  and $\beta_m=0.6$.}}
  \label{five}
\end{figure}

Figure \ref{three} demonstrates how the SUs' sum rate is affected by the cascade level $(\kappa)$, the number of SU transmitter antennas $(A)$, and the number of SU receivers $(B)$. We observe that an augmentation in $\kappa$ leads to a deterioration in the communication quality of the SUs. The reason for this is because a larger value of $\kappa$ corresponds to a greater number of objects blocking the paths, resulting in more prominent fading and hence less reliability of the connections. Furthermore, it is evident that as the number of SU receivers rises, the sum rate decreases. This occurs because each SU receiver not only receives its own signals, yet also the messages of other SUs, resulting in an escalation of inter-user interference.

 Figure \ref{seven} illustrates the average reward of the SUs associated with various values of $\beta_m$ and $P_t$. A higher value of $\beta_m$ indicates that the components of the RIS have a greater ability to reflect a larger amount of signal power, regardless of the phase shift applied. This enhances the signal intensity received by the SU receiver, leading to an increased data rate.   We observe that the transmission quality of the SUs improves as the transmit power increases, mainly because the received   SINR at the SU receivers is enhanced. Moreover, this figure demonstrates a  trade-off between power efficiency and RIS efficiency. Specifically, when the RIS components have a lower power reflection capability, namely when $\beta_m=0.2$, the effect of increasing $P_t$ from $1$ dB to $30$ dB is greater than when $\beta_m=0.9$, which is the point at which most of the power of the received signal is transmitted to the SU receivers. This implies that in situations where the RIS is very efficient, the investment in additional power has a minimum effect, enabling possible reductions in power consumption while yet ensuring satisfactory performance. Conversely, when the value of $\beta_m=0.2$, the SUs depend more on enhancing the transmission power to address the inefficiencies in the RIS hardware. Moreover, taking into account the hardware limitations regarding imperfect reflection offers a more accurate evaluation of the performance, rendering the implemented solutions robust and effective for real-world applications where achieving perfect reflection, i.e., when $\beta_m=1$,  is not possible.

{\color{black} \subsection{A Comparison between the SAC and other DRL approaches}}
{\color{black}To illustrate the superiority of the proposed SAC approach, we compare it with two DRL methodologies and a Random policy as depicted in Figure \ref{five}.   Additionally, we analyze the impact of increasing the number of PUs on the SUs rate in Figure \ref{six}.}

{\color{black} Figure \ref{five} presents a comparison of the proposed SAC technique against two DRL benchmarks: TD3 and DDPG. This comparison has been conducted as hybrid RIS-assisted CRNs have not yet been extensively explored in the literature. DDPG is a DRL technique recognized for its efficiency in continuous state and action spaces, which applies to our scenario.   TD3 enhances the stability of DDPG by mitigating overestimation bias through the use of twin critics and delayed policy updates, making it more robust for complex continuous control tasks. On the other hand, in a random strategy, the agent randomly chooses the beamforming matrix   and   the RIS parameters, without acquiring knowledge from the environment. 
We can notice that the SAC method outperforms TD3 and  DDPG, demonstrating a superior level of stable convergence and higher cumulative rewards. This improvement can be attributed to SAC’s entropy regularization mechanism, which promotes efficient exploration and prevents premature convergence to sub-optimal policies. Moreover, all DRL procedures surpass the random policy, as the latter selects actions arbitrarily, whereas the agent in DRL methodologies learns optimal actions through environmental interaction and receives rewards or penalties to enhance decision-making over time.}
 
   \begin{figure}  [t] 
  \centering
  \includegraphics[width=1.0\linewidth]{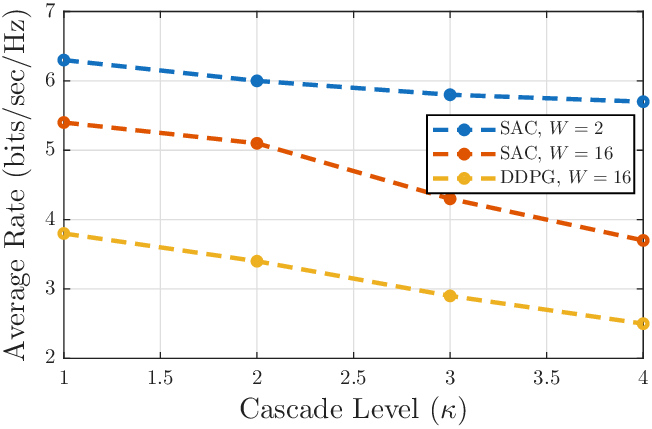}
  \caption{SUs' average reward (rate) for different number of PU receivers $(W)$ for the SAC and DDPG policies.  $P_{t}=30$ dB, $\kappa_s=\kappa_b=\kappa$, and $I=10$ dB.}
  \label{six}
\end{figure}


Figure \ref{six} displays the average reward of the SUs as a function of the cascade level, considering various numbers of PU receivers ($W$), using the SAC and DDPG methods. As the value of $W$ grows, the average reward  decreases.  This behaviour may be attributed to the interference restriction imposed by the PU receivers, as stated in equation (\ref{const}). This restriction guarantees that the transmit power of the SUs is restricted by the interference threshold that is the most restrictive among all PU receivers. As $W$ grows, the probability of meeting a PU receiver with a stronger channel characteristics, i.e., higher channel  power gain $(g_{spi})$, for $i \in\{1,2,\cdots,W\}$, also rises. Consequently, there is a more stringent interference limitation, which ultimately decreases the permissible transmission power of the SUs.

{\color{black} \subsection{Impact of the energy threshold $(\tau)$ on the hybrid RIS, and  a comparison between passive, active, and hybrid RIS (dynamic and fixed)}}

{\color{black} To illustrate the impact of the proposed dynamic hybrid RIS, we analyze the impact of the minimum energy required for the RIS to operate in active mode $(\tau)$ as depicted in Figure \ref{mix} and Figure \ref{chart}.  This subsection also presents a comparison of passive, active, and hybrid RIS.  We conclude with a comparison of the dynamic hybrid RIS with the fixed hybrid RIS employed in previous studies in Figure \ref{fix_hybrid}.}



 \begin{figure}  [t] 
  \centering
  \includegraphics[width=1.0\linewidth]{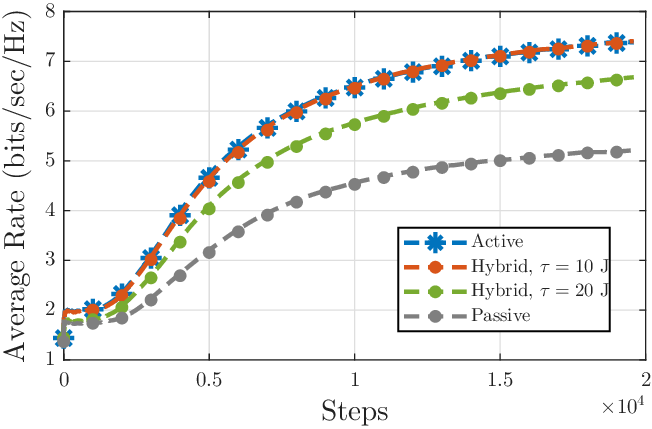}
  \caption{A Comparison of SUs average rate for passive, active, and dynamic hybrid RIS at varying  $\tau$. }
  \label{mix}
\end{figure}

   \begin{figure}  [t] 
  \centering
  \includegraphics[width=0.9\linewidth]{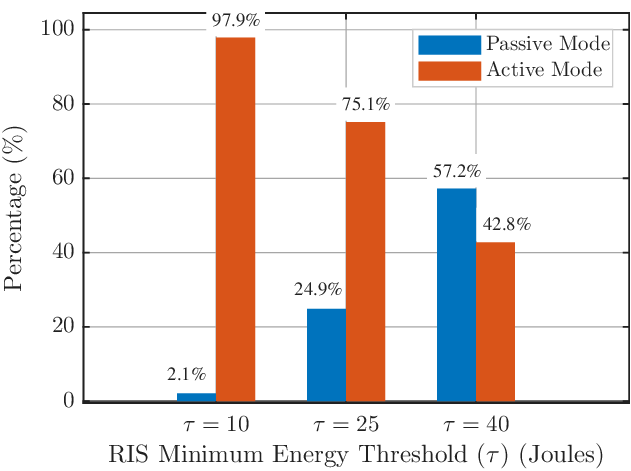}
  \caption{A comparison between  the passive and active mode usage in hybrid RIS system.}
  \label{chart}
\end{figure}

 Figure \ref{mix} depicts a comparison among fully passive RIS, fully active RIS, and dynamic hybrid RIS for two values of the minimal energy required to operate in the active mode $(\tau)$.  It is noted that when $\tau$ increases, the agent's reward degrades due to the increased level of required energy to operate in the active mode, resulting in a greater likelihood that the RIS will function in passive mode for the majority of the steps.  In this instance, the link's reliability will be degraded as the passive mode does not enhance the signal's power; it simply steers the signals to the destination. Moreover, the average SUs' rate is the lowest in the passive scenario due to the lack of signal amplification.  With a reduced threshold, namely for $\tau=10$, the hybrid mode has the same performance as the active mode, since the RIS operates more frequently in the active mode due to the reduced energy requirements, while still conserving energy when feasible.  This intentional switching allows the hybrid method to leverage the amplification advantages of active RIS without facing its continuous energy expenses.


 Figure \ref{chart} demonstrates the frequency with which the hybrid RIS operates in either active or passive mode, serving as an indicator of the system's energy-aware behavior.  We conclude that with an elevated threshold $\tau$, the agent favors passive mode over active mode to prevent consuming energy for amplification that surpasses the harvested energy, signifying a balance between performance and energy efficiency.  For example, for $\tau=10$ Joules, the proposed structure operates as an active RIS for $97.9\%$ of the time, indicating frequent amplification and increased energy consumption.  Nevertheless, a more balanced mode-switching, illustrated by $\tau=40$ Joules, where the RIS is $42.8\%$ of the time active and $57.2\%$ is passive, underscores the RIS's capacity to preserve energy while optimizing the reliability of the SUs.    The clear different operation mode demonstrates how energy limitations directly affect the RIS operational strategy, and energy saving must be a critical consideration when implementing the active RIS.  The suggested hybrid RIS architecture is particularly attractive for practical implementation in energy-limited wireless systems, such as cognitive IoT networks, where it is essential to balance performance and power consumption.

  \begin{figure}  [th] 
  \centering
  \includegraphics[width=0.9\linewidth]{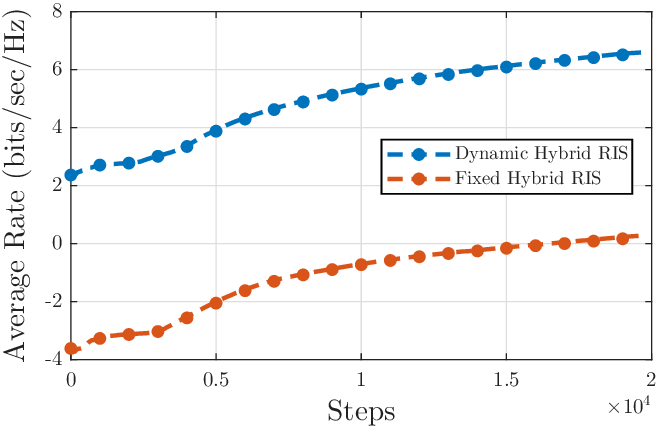}
  \caption{A comparison between the dynamic hybrid RIS and fixed hybrid RIS. $\tau=20$ Joules, $P_t=30$ dB, $I=20$ dB, and $\kappa_s=\kappa_b=2$.    }
  \label{fix_hybrid}
\end{figure}

Figure \ref{fix_hybrid} illustrates a comparison between the proposed dynamic hybrid RIS and the fixed hybrid RIS. For the fixed hybrid mode, such as the one deployed in \cite{10483088,10810415,10807185},    $50\%$ of the RIS elements are designated as active and amplifying with a fixed gain of $2$, while the remaining elements are passive.  The proposed dynamic approach surpasses the fixed hybrid RIS by adaptively activating elements when sufficient power is available, hence improving the data rate.  The fixed hybrid RIS will consistently utilize half of the elements as passive, despite the potential to activate them for amplification when sufficient energy is available.  This lowers the agent's reward and consequently the link's data rate.

{\color{black} \subsection{Energy consumption in active and hybrid RIS}
This section illustrates the energy consumption associated with both active and hybrid modes in Table \ref{table1}.  Furthermore, we present a trade-off between performance and energy efficiency offered by the dynamic hybrid RIS in Figure \ref{scatter}.}

Table \ref{table1} illustrates a comparison of the energy consumption in Joules (J) between active and hybrid modes for various values of $\tau$. For the active RIS, the power consumed per unit amplification and the control circuit power per element are given respectively as $P_{\text{amp}}=50$ mW, $P_{\text{ctrl}}=10$ mW, while the  per-element control power to operate the passive RIS is set to $P_{\text{passive}}=0.1$ mW.  The energy usage is contingent upon the minimal threshold $\tau$ of the hybrid RIS.  Specifically, lower values of $\tau$ result in increased energy consumption by the RIS, as it operates more frequently in the active mode rather than in the passive mode, which consumes a negligible amount of energy to operate.  In the fully active RIS, the energy consumption is nearly invariant with respect to $\tau$ and continuously remains high across all $\tau$ levels.  The hybrid RIS not only bridges the performance gap between active and passive architectures but also delivers substantial energy savings. 

{\color{black}
Figure \ref{scatter} illustrates the performance–energy   trade-off of the proposed hybrid RIS system compared with the fully active RIS.   As $\tau$ increases, the RIS operates in the passive mode more frequently, reducing  energy consumption but lowering the SU rate. Conversely, smaller $\tau$  values yield higher rates at the expense of greater energy use.  The active RIS (orange square)   serves as an upper performance bound but with a consistently higher energy cost.  
This visualization   enables network designers to select an appropriate $\tau$  depending on system priorities (e.g., energy-constrained IoT vs. throughput-oriented indoor networks). The parameter  $\tau$ controls the switching sensitivity between the passive and active RIS states. Its optimal value depends on the harvested energy availability: smaller $\tau$ values are preferable in dense or high-interference settings, to preserve performance, while larger $\tau$ values are better suited  for sparse or energy-limited scenarios (e.g., remote IoT nodes) to improve energy efficiency. Thus,  $\tau$ serves as a tunable system-level knob that can be adapted to topology dynamics or power-budget constraints to balance performance and energy.}


 \begin{table}[t]
\centering
\caption{Average Energy Consumption Comparison. }
\label{table1}
\begin{tabular}{|c|c|c|c|}
\hline
\textbf{$\tau$ (J)} & \textbf{Hybrid (J)} & \textbf{Active (J)} & \textbf{Energy Savings (\%)} \\
\hline
50 & 0.092 & 0.356 & 74.2\% \\
40 & 0.148 & 0.343 & 56.8\% \\
30 & 0.226 & 0.354 & 36.3\% \\
10 & 0.306 & 0.352 & 13.1\% \\
\hline
\end{tabular}
\label{table1}
\end{table}

  \begin{figure}  
  \centering
  \includegraphics[width=0.9\linewidth]{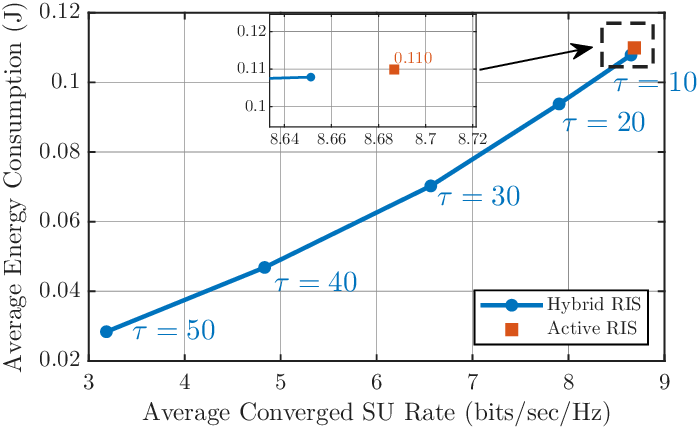}
  \caption{{\color{black}A trade-off between energy consumption and performance (average converged rate) for the fully active and hybrid RIS for different values of $\tau$.   $P_t=30$ dB, $I=10$ dB, and $\kappa_s=\kappa_b=2$.  }  }
  \label{scatter}
\end{figure}

 \begin{figure} [t]   
  \centering
  \includegraphics[width=1\linewidth]{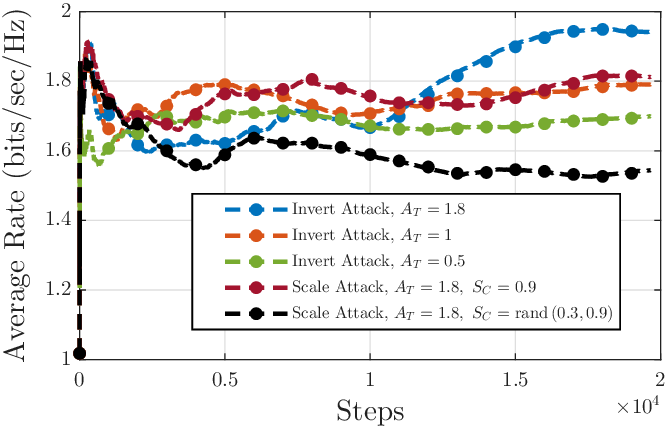}
  \caption{{\color{black} The SUs' average rate for the invert and scale poisoning attacks for different values of the reward threshold $(A_T)$ and the scaling factor $(S_C)$. $\tau=30$ Joules, $P_t=30$ dB, $I=20$ dB,  and $\kappa_s=\kappa_b=2$.}}
  \label{invert_and_scale}
\end{figure}


{\color{black} \subsection{Impact of the proposed attack and defense}}

{\color{black} This section analyzes the impact of inversion and scale reward poisoning attacks on the SU's rate and the agent's learning, as seen in Figure \ref{invert_and_scale}.   We also illustrate the effect of the reward threshold $(A_T)$.   Additionally, we highlight the impact of the proposed defense mechanism on the agent's learning process by comparing the average reward with and without the defense, as depicted in Figure \ref{attack_and_defense}.}

 \begin{figure}  [t]  
   \centering
  \includegraphics[width=1.0\linewidth]{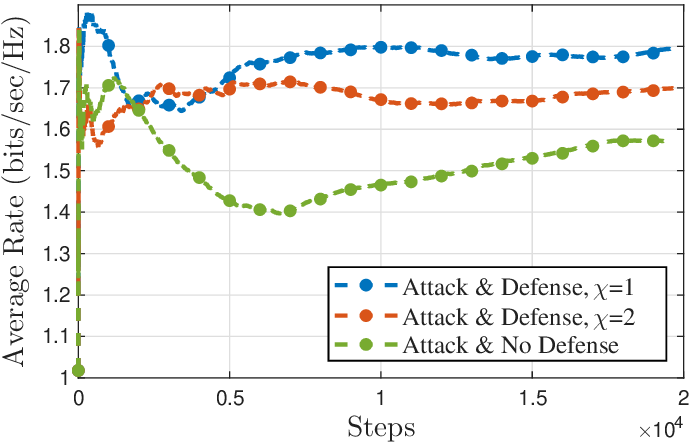}
  \caption{ {\color{black}Impact of the defense mechanism  on the invert poisoning attack. $\tau=30$ Joules, $P_t=30$ dB, $I=20$ dB, $A_T=0.5$, and $\kappa_s=\kappa_b=2$.}}
  \label{attack_and_defense}
\end{figure}

{\color{black}
The invert and scale reward poisoning attacks, as well as the influence of the reward threshold $(A_T)$ and the scaling parameter $(S_C)$ on the agent's average reward, are illustrated in Figure \ref{invert_and_scale}.  We discovered that the performance of the DRL agent improves as $A_T$ increases, as the attacker will perform attacks less frequently since Q-values must be quite high.  It is intriguing to observe that for $A_T=1.8$, the scale attack results in a more effective suppression of the agent's learning as opposed to the invert attack.   This is due to the fact that it consistently reduces the value of rewards, which prevents the agent from reinforcing positive actions over time, resulting in poor performance. {\color{black} To further examine the impact of attack intensity, we varied both the reward threshold and the scaling factor to emulate weak and strong adversarial conditions. In particular, the random-scale attack—where the scaling factor $S_C$ is uniformly sampled from $(0.3, 0.9)$—represents a variable-intensity scenario that exposes the agent to fluctuating levels of distortion.} That is,  although the random attack leads to lower and more fluctuating performance compared to fixed-scale attacks, the agent retains a stable learning trajectory, highlighting resilience to stochastic adversarial behavior.

Figure \ref{attack_and_defense} depicts the impact of the proposed defense mechanism  on the considered invert poisoning attack.  We notice that the proposed defense mechanism significantly improves the agent's resilience to the   reward poisoning attack. While the attack without defense causes the agent to converge to a suboptimal policy, the defense helps maintain a stable and higher average rate over time. {\color{black} Moreover, conceptually, the filtering threshold $\chi$  controls the trade-off between robustness and performance; a smaller $\chi$ applies stricter rejection of poisoned rewards, potentially limiting exploration in benign settings. However, as shown in the figure, the defense not only improves robustness but also yields higher average rates across all steps indicating that with the tested attack severity, tightening the filter ($\chi=1$) effectively removes corrupted rewards without penalizing normal learning.} This highlights the effectiveness of combining reward clipping and outlier filtering based on historical statistics in mitigating adversarial interference.}

\section {Scalability, Complexity and Security Generalization}
While our SAC-based solution proves highly effective for RIS-CRN scenarios of moderate size, future deployments—such as those involving much larger RIS surfaces or massive-multiple-input multiple-output (MIMO) arrays—will benefit from additional strategies to maintain scalability and real-time responsiveness. Potential pathways include hierarchical or distributed control which could help manage the increased computational load without compromising performance.
Looking ahead, as these networks continue to evolve and attract wider adoption, it will be beneficial to also examine other possible security risks. {\color{black}  While our current defense is tailored to reward-driven attacks, the same principles of lightweight statistical monitoring may be adaptable to broader attack scenarios.  Moreover, since  the defense relies on detecting statistical deviations rather than attack-specific patterns, it can handle diverse manipulations such as biased shifts, spikes, bursts, or stochastic perturbations in the reward stream, rendering the proposed defense technique applicable to a variety of reward poisoning attacks on DRL approaches.} Exploring such extensions and developing unified approaches to enhance robustness against a diverse set of adversarial strategies represents an interesting direction for future work.

\section{Conclusions}
This article examines the application of   dynamic hybrid reconfigurable intelligent surfaces (RISs) in an underlay cognitive radio network, with the presence of multiple primary users. A deep reinforcement learning (DRL) technique was used to maximize the downlink rate of the secondary users. This was achieved by optimizing both the transmit beamforming and the reflection phases of the RIS. The results demonstrated the agent's capacity to achieve optimal long-term rewards through the convergence of the results. Furthermore, we demonstrated that the SAC method surpasses the DDPG, TD3, and random approaches. In addition, we showed that augmenting the number of elements within the RIS and increasing the value of the minimum reflection amplitude directly enhance the rate of the secondary users. Moreover, our findings suggest that increasing the cascade level of the channels and the number of SU receivers has a detrimental effect on the reliability of the users. Furthermore, we demonstrated that the hybrid RIS surpasses the passive RIS in performance. Moreover, our suggested model  consumes less energy compared to active RIS.  That is, the proposed model balances performance and energy efficiency. Our results also showed that the proposed dynamic hybrid RIS outperforms the fixed hybrid RIS. {\color{black} Finally, we showed that  our proposed defense mechanism significantly mitigates the impact of reward poisoning attacks, maintaining a stable average rate performance even under adversarial conditions. By rigorously modeling the interplay between energy harvesting, dynamic RIS configuration, and adversarial learning, our work takes a vital step toward practical, secure, and efficient cognitive radio systems. Deploying these methods in hardware is a key direction, requiring further study of control granularity, energy-delay tradeoffs, and secure feedback mechanisms.}


\end{document}